\documentclass[12pt]{iopart}

\newcommand{\aver}[1]{\langle{#1}\rangle}
\newcommand{\trace}[1]{\tr({#1})}

\newcommand{\eps}{\varepsilon}
\newcommand{\pd}{\partial}
\newcommand{\bx}{{\bf x}}
\newcommand{\bfr}{{\bf r}}
\newcommand{\bk}{{\bf k}}
\newcommand{\bv}{{\bf v}}
\newcommand{\im}{{\rm i}}

\begin{document}

\title[Anomalous scaling in two and three dimensions for
a passive vector advection]{Anomalous scaling in two and three
dimensions for a passive vector field advected by a turbulent
flow}

\author{S V Novikov}
\address{
    Department of Theoretical Physics, St.~Petersburg University,
    Ulyanovskaya 1, St.~Petersburg, Petrodvorets, 198504 Russia }
\ead{snov@mail.ru}

\begin{abstract}
A model of the passive vector field advected by the uncorrelated
in time Gaussian velocity with power-like covariance is studied by
means of the renormalization group and the operator product
expansion. The structure functions of the admixture demonstrate
essential power-like dependence on the external scale in the
inertial range (the case of an anomalous scaling). The method of
finding of independent tensor invariants in the cases of two and
three dimensions is proposed to eliminate linear dependencies
between the operators entering into the operator product
expansions of the structure functions. The constructed operator
bases, which include the powers of the dissipation operator and
the enstrophy operator, provide the possibility to calculate the
exponents of the anomalous scaling.
\end{abstract}

\section*{Introduction}
The term ``anomalous scaling'' refers to deviations from the
predictions of the Kolmogorov theory. Such deviations take place,
in particular, in the behaviour of the structure functions
$S_n(r)$ of the turbulent velocity field $\bv$ in the inertial
range $ l\ll r \ll L $, where $L$ and $l$ are the external scale
and the dissipation length. The structure functions of the vector
field $\varphi$ are defined as
\begin{equation}\label{str}
S_n(r)\equiv\aver{[\varphi_r(t,\bx+\bfr)-\varphi_r(t,\bx)]^n},
\qquad \varphi_r \equiv \varphi_i r_i/r,
\end{equation}
For the field $\bv$ the Kolmogorov theory predicts $S_n(r)\sim
r^{n/3}$, while the experiments indicate that $S_n(r)\sim
r^{n/3-\xi_{n}}$ with certain nontrivial ``anomalous exponents''
$\xi_{n}$.

The Kolmogorov theory states that the only dimensional parameter
which determines the statistics of the turbulent velocity
pulsations is the average rate of energy dissipation per unit mass
$\overline{\varepsilon}=\nu \aver{\Phi_{\rm dis} }$, where $\nu$
is the kinematic viscosity and $\Phi_{\rm dis}\equiv(\pd_i
v_j+\pd_j v_i)^2/2 $ is the operator of local energy dissipation.
In stationary state, it is held that $\overline{\varepsilon}=W$
with the power of energy pumping $W$. In this case the Kolmogorov
hypothesis can be formulated for the structure functions
(\ref{str}) as follows. Dimensional analysis leads to the relation
$S_n(r,\nu,W,L)=(W r)^{n/3}R_n(r/l,r/L) $, where $l\equiv
(\nu^3/W)^{1/4}$. Then the Kolmogorov hypothesis states that
function $R_n$ has finite limit in the inertial interval:
$R_n(\infty,0)=const\neq 0$.

At present the possibility of taking limit in the $R_n(r/l,r/L)$
on the first argument is universally recognized, i.e. the quantity
$R_n(r/L)\equiv R_n(\infty,r/L)$ is recognized finite. Hence the
structure functions are considered to be independent on the
viscosity in the IR region $r\gg l$. If the Kolmogorov hypothesis
about finiteness of the $R_n(r/L)$ in the limit $r/L\to 0$  is
violated and the asymptotic behaviour is power-like $R_n(r/L)\sim
(r/L)^{-\xi_n}$ then the anomalous scaling appears. In
phenomenological generalizations of the Kolmogorov theory the
anomalous scaling is usually considered as a result of
fluctuations of the energy dissipation rate \cite{Monin,Frisch}.

The analogy with the theory of critical phenomena suggests itself
when the problem is formulated as the field-theoretical model.
Then the direct analogy to the theory of critical phenomena
provides possibility to apply the powerful UV-renormalization
technique to the problem. In this way the first Kolmogorov
hypothesis can be proved (independence of the structure functions
on the viscosity in the inertial range). The behaviour of
$R_n(r/L)$ in the range $r/L\ll1$ is not determined by the
renormalization group itself, the method to find it is the
operator product expansion (OPE) which gives the asymptotic
expansion
\begin{equation}\label{intro:Snr}
  S_n(r)\propto r^{n/3} \sum\nolimits_F
  (r/L)^{\Delta_F}A_F,
\end{equation}
where summation runs over all possible scalar operator products
$F$ (constructed from local products of the fields $v_i(t,\bx)$
and their derivatives), $\Delta_F$ are the critical dimensions of
the operators
 \cite{2loop
}. More precisely, $\Delta_F$ are the eigenvalues of the matrix of
critical dimensions, and summation in (\ref{intro:Snr}) runs over
the eigenvectors of the matrix. The operators entering into the
OPE are those which appear in the Taylor expansion and all the
operators that admix to them in renormalization.

If for each $F$ holds $\Delta_F>0$ (as in the theory of phase
transitions) then the terms in (\ref{intro:Snr}) determines
corrections to the Kolmogorov scaling. If there is an operator in
(\ref{intro:Snr}) with $\Delta_F<0$ --- ``dangerous operator''
then the limit $r/L\to 0$ in (\ref{intro:Snr}) does not exist
which leads to the anomalous scaling. Realization of the described
scenario meets however two difficulties: technical difficulty of
calculation of the critical dimensions $\Delta_F$ and fundamental
one --- if in the theory a dangerous operator exists then it can
be shown that with necessity there are infinitely many dangerous
operators, the spectrum of their critical dimensions is unbounded
below (we can not point at ``the most dangerous'' operator). Thus
the expansion (\ref{intro:Snr}) can be useful when summation of
the series is possible, or when the (\ref{intro:Snr}) contains
only finite number of terms due to the model features.

\section{A passive scalar admixture}
Recently, significant progress in the description of the anomalous
scaling has been achieved in related problems of the turbulent
advection of a passive admixture. The experiments and computer
simulation data demonstrate that the anomalous scaling appears not
only in the turbulent pulsations of the velocity, but much more it
reveals in the properties of the field transferred by the
turbulent flow $\theta(t,\bx)$ which may be the field of the
admixture concentration or the temperature field. The passive
scalar advected by the turbulent velocity field $v_i(t,\bx)$ is
described by equation
\begin{equation}\label{model:dynamix}
  \pd_t\theta+(v_j\pd_j)\theta=\nu_0\pd^2\theta+f,
\end{equation}
where $\nu_0$ is the diffusivity (or the thermal diffusivity), $f$
is the source of the field $\theta$. Significant progress in the
description of the anomalous scaling has been achieved in a
relatively simple model, due to Kraichnan \cite{Kraich1,Kraich2},
of a passive scalar advection which assumes the velocity field to
be delta-correlated in time and Gaussian with covariance
\begin{equation}\label{corr_v}
  \aver{v_i(t+\tau,\bx+\bfr)v_j(t,\bx)}=
  D_0\,\delta(\tau)\int\frac{d\bk}{(2\pi)^d}P_{ij}(\bk)
  N(k)\exp\im\bk\bfr,
\end{equation}
where $P_{ij}({\bk})\equiv \delta_{ij}-k_i k_j/k^2 $ is the
transverse projector (the consequence of transversality of the
velocity), the function $N(k)$ in $d$-dimensional space is
modelled with power-like expression $N(k)\equiv k^{-d-\eps}$ and
is supposed to be somehow IR regularized on the scale $k\sim
L^{-1}$.
 Here $0 \leq \eps \leq 2$ is a kind of H\"older exponent which
measures the ``roughness'' of the velocity field. In the RG
approach, it plays the same role as the parameter $4-d$ in the
theory of critical phenomena.
 The physical value of
parameter $\eps$ is $\eps=4/3$, for which the amplitude factor
$D_0$ has the same dimension as the energy pumping power per unit
mass.

In the model (\ref{model:dynamix})--(\ref{corr_v}), the existence
of the anomalous scaling was substantiated and the exponents
$\xi_n$ of the structure functions were calculated
\cite{GK,BGK,Falk1,Falk2}. The first results were obtained using
the Hopf equations on the distribution function of the equal time
fluctuations of the admixture field. The linearity of the equation
(\ref{model:dynamix}) for $\theta$ results in closed equations
(instead of chain of equations) for the equal time correlation
functions. The analysis of so-called ``zero modes'' allows to
substantiate the anomalous scaling and to calculate the anomalous
exponents. The pair function is founded exactly and has no
anomalous scaling ($\xi_2=0$). The exponents of the high-order
functions were calculated approximately with small parameters
$\eps$ \cite{GK,BGK} and $1/d$ \cite{Falk1,Falk2}:
\begin{equation}\label{dz1e}
\xi_n=n(n - 2)\eps /2(d + 2) + O(\eps^2),
\end{equation}
\begin{equation}\label{dz1d}
\xi_n=n(n - 2)\eps /2d + O(1/d^2).
\end{equation}
The analysis of the zero modes gives also interesting results in
another case --- in the Batchelor limit $\eps=2$
\cite{Bal,Pumir1,Pumir2,Pumir3}.

Until now the technical difficulties do not allow to calculate
higher-order terms of the expansions (\ref{dz1e}) using analysis
of the zero modes. The methods of RG and OPE turns out to be more
effective to find the asymptotic approximation on small $\eps$.
The terms $\sim \eps^2$ \cite{2loop} and $\sim \eps^3$
\cite{3loop} were calculated using these methods. It is essential
that in the Kraichnan model the finite number of terms contribute
to the OPE (\ref{intro:Snr}) of the structure function $S_n$,
precisely, those are the powers of the dissipation operator of
admixture field $\Phi_{\rm dis}^k$ with $1 \leq k \leq n/2$.
Joining of asymptotic expansion of the exponents $\xi_n$ on small
$\eps$ (taking three terms of the expansion) with asymptotic
approximation near the Batchelor limit($\eps=2$) results in
interpolation for $\xi_n(\eps)$ which coincides with the data of
numerical experiment (within the experimental errors) in the range
$0 \leq \eps \leq 2$, including the physical value $\eps=4/3$
\cite{3loop}.

The RG method demonstrates the universality of the approach
besides technical efficiency. Using the RG the anomalous exponents
of order $\eps^2$ in the model of compressible fluid were
calculated \cite{RG1}, the model of advection by the Gaussian
velocity field with finite correlation time was analyzed in
\cite{RG3,Juha2}, the Kazantsev model of the magnetic
hydrodynamics --- in \cite{RG1}.

\section{A passive vector admixture}
In the present paper we concentrate on the model of advection of a
passive vector admixture proposed in
\cite{Runov,Procaccia,Yap
,TMF03}. The model is the direct generalization of
(\ref{model:dynamix}) and is described by the equation
\begin{equation}\label{model:dynamix1}
  \pd_t\varphi_i+(v_j\pd_j)\varphi_i=\nu_0\pd^2\varphi_i-\pd_i P+f_i,
\end{equation}
which come out from equation (\ref{model:dynamix}) after replacing
of the scalar field $\theta$ with the transversal vector field
$\varphi_i$ and adding of the pressure gradient $P$ to the left
hand side to ensure transversality $\pd_i\varphi_i=0$. The random
force $f_i$ in (\ref{model:dynamix1}) obeys the Gaussian
distribution with zero mean and the covariance
\begin{equation}\label{model:corr_f}
  \aver{f_i(t+\tau,\bx+\bfr) f_j(t,\bx)}=\delta(\tau)\,C_{ij}(\bfr/L).
\end{equation}
The exact form of the function $C_{ij}(\bfr/L)$ is not important.
It only models the energy influx resulting from the interaction
with large vortices, which compensates the dissipation losses.

Note that all the odd-order structure functions vanish because the
equation (\ref{model:dynamix1}) is invariant with respect to the
substitution $\varphi \to -\varphi$, $f \to -f$ and the fact that
$f$ is Gaussian (the pressure term can be eliminated from
(\ref{model:dynamix1}) by applying the transverse projector).

The model (\ref{corr_v}), (\ref{model:dynamix1}),
(\ref{model:corr_f}) can be considered as a rough approximation
describing the turbulent velocity field, when $\varphi$ is
regarded as the ``hard'' component of the velocity field and $v$
is regarded as the ``soft'' one. Then in the Navier--Stokes (NS)
equation for the hard component only the convective transfer by
the soft component is taken into account and the soft component
obeys Gaussian distribution with covariance (\ref{corr_v}). These
assumptions do not violate Galilean invariance of the model. Since
the structure functions (\ref{str}) are Galilean-invariant, it
follows that OPE (\ref{intro:Snr}) contains only contributions
from the operators which are also Galilean-invariant, just as in
the stochastic hydrodynamics which is described by the NS
equation.

The model  (\ref{corr_v}), (\ref{model:dynamix1}),
(\ref{model:corr_f}) was analyzed by means of the Hopf equation
for the pair correlation function \cite{Procaccia}. The function
turns out to be non-anomalous ($\xi_2=0$) in the isotropic case.
The model was considered using the methods of RG and OPE in
\cite{Runov}. It was shown, that the main contribution to the OPE
of the $S_{2n}$ gives a family of scalar operators of the form
$(\pd\varphi)^{2n}$. Renormalization of the operators is not
multiplicative, but includes operator mixing. The mixing
complicates greatly calculation of the anomalous exponents of the
operators; obtaining of an analytic expression of the anomalous
exponents for arbitrary $n$ do not succeed even in one-loop
approximation. In \cite{Runov}, the one-loop approximation was
considered for the family of the $S_4$; the exponents were
evaluated in the linear on $\eps$ approximation, the negative
exponent were found among them, so the existence of the anomalous
scaling was proved in the model of the vector advection with the
mixing of the operators. It was in the case of flat flows ($d=2$)
\cite{TMF03} that the anomalous exponents of the structure
functions of the arbitrary order were calculated in the model
(\ref{corr_v}), (\ref{model:dynamix1}), (\ref{model:corr_f}). The
anomalous exponents were obtained in the one-loop approximation
coincide with the exponents (\ref{dz1e}) of the scalar admixture.
``The most dangerous'' operators were found to be the powers of
dissipation operator, just as in the scalar model.

\section{The bases for scalar operators of the form $(\pd\varphi)^n$}
As demonstrated in \cite{Runov}, the leading terms in
(\ref{intro:Snr}) arise from the family $\Phi^{(n)}$ of scalar
operators of the form $(\pd\varphi)^{n}$. More precisely, that
family consists of all possible contractions of $n$ tensors
$\pd_i\varphi_j$. Denoting the symmetric and the antisymmetric
parts of the $\pd_i\varphi_j$ as
\begin{equation}\label{notation}
    S_{ij}\equiv(\pd_i\varphi_j+\pd_j\varphi_i)/2,\qquad
    A_{ij}\equiv(\pd_i\varphi_j-\pd_j\varphi_i)/2,
\end{equation}
we can express for $S_4$ the seven operators which forms the
family $\Phi^{(4)} = $ $\{\trace{A^2}^2$, $\trace{S^2}^2$,
$\trace{A^2}\trace{S^2}$, $ \trace{A^2 S^2}$, $ \trace{(AS)^2}$, $
\trace{A^4}$, $\trace{S^4}\}$. The main difficulty in the
calculation of the critical dimensions of the operators from
$\Phi^{(n)}$ is the rapid growth with $n$ of the number of
relevant operators. For example, the family $\Phi^{(6)}$ contains
$24$ operators, $\Phi^{(8)}$ --- $81$, $\Phi^{(10)}$ --- $278$,
and $\Phi^{(18)}$ --- as many as $47246$ operators. Fortunately,
not all of them are independent for sufficiently small space
dimensions $d$. For the most interesting cases $d=2$ and $d=3$,
the elimination of redundant operators reduces drastically the
size of the matrix of critical dimensions.

Our goal is to construct for $d=2$ and $d=3$ the sets $\Omega_d$
of minimal size which contains invariants of the matrix
$\pd_i\varphi_j$. Then we will form bases of $\Phi^{(n)}$ using
products of the invariants from $\Omega_d$.

\subsection{2-d case} The dependencies mentioned above can be
considered as consequences of the Hamilton--Cayley identity. For a
traceless $2\times2$ matrix $M$ it has the form:
\begin{equation}\label{2x2:chiA}
  \chi_2(M)=M^2 - I \trace{M^2}/2=0,
\end{equation}
with identity matrix $I$. From (\ref{2x2:chiA}) one has
$\chi_2(\alpha A+\beta S)=0$ for any linear combination of the
matrices $A$ and $S$ from (\ref{notation}). Collecting the
coefficients of three independent structures $\alpha^{2}$,
$\beta^{2}$ and $\alpha\beta$ gives three relations:
\begin{equation}\label{2x2:=}
  A^2- I\trace{A^2}/2=0,\quad
  S^2- I\trace{S^2}/2=0,\quad
  AS+SA =0.
\end{equation}
The first and the second relations are the Hamilton--Cayley
identities for the matrices $A$ and $S$. The last one can be
interpreted as a commutation relation. The relations (\ref{2x2:=})
can be written in the form
\begin{equation}\label{2x2:sim}
  A^2 \sim 0, \quad S^2 \sim 0, \quad SA \sim -AS,
\end{equation}
where $\sim$ means the equality up to polynomials of lesser
degree.

It follows from (\ref{2x2:sim}) that $P_3(A,S) \sim 0$ for any
polynomial of the third degree, so that an arbitrary polynomial of
the matrices $A$ and $S$ can be written as $P(A,S) = C_1(\Omega_2)
AS +C_2(\Omega_2) A +C_3(\Omega_2) S+ C_4(\Omega_2) I$, where the
coefficients are polynomials of the scalar invariants from the
minimal set
\begin{equation}\label{Omega2}
    \Omega_2(A,S)=\{ \trace{A^2}, \trace{S^2}\}
\end{equation}
and, as a consequence, $\trace{P(A,S)}=2 C_4(\Omega_2)$.

Thus for $d=2$ each operator from $\Phi^{(2n)}$ is a polynomial
$P_n(\trace{A^2}$, $\trace{S^2})$, which can be decomposed in the
basis
\begin{equation}\label{basis2}
  \trace{A^2}^{n-k}\trace{S^2}^k,\quad 0\leq k \leq n.
\end{equation}

\subsection{3-d case} The Hamilton--Cayley identity for a
traceless $3\times3$ matrix $M$ has the form:
\begin{equation}\label{3x3:chiA}
  \chi_3(M)= M^3 - M \trace{M^2}/2-I\trace{M^3}/3=0.
\end{equation}
From the Hamilton--Cayley identity for the matrices $A$ and $S$
from (\ref{notation}) it follows that the minimal set
$\Omega_3(A,S)$ contains the invariants $\trace{A^2}$, $
\trace{S^2}$ and $ \trace{S^3}$. Equating to zero the coefficients
of $\alpha^2 \beta$ and $\alpha \beta^2$ in the identity
$\chi_3(\alpha A+\beta S)=0$
 gives:
\begin{equation}\label{3x3:chiAAS}
  A^2S + ASA + SA^2 - S\trace{A^2}/2  - I\trace{A^2S} = 0.
\end{equation}
\begin{equation}\label{3x3:chiASS}
  AS^2 + SAS + S^2A - A\trace{S^2}/2  = 0.
\end{equation}
It follows from (\ref{3x3:chiAAS}) that the invariant
$\trace{A^2S}$ is in the minimal set $\Omega_3(A,S)$. Neglecting
the polynomials of lesser degrees in
(\ref{3x3:chiA})--(\ref{3x3:chiASS}) results in commutation-like
relations
\begin{equation}\label{3x3:simAAA}
  A^3 \sim 0, \quad  S^3 \sim 0,
\end{equation}
\begin{equation}\label{3x3:simSSA}
  SAA \sim -AAS -ASA,\quad SSA \sim -ASS -SAS.
\end{equation}
It follows from (\ref{3x3:simAAA}) that each monomial containing
$A^3$ or $S^3$ vanishes. Each monomial containing $SAA$ or $SSA$
can be decomposed according to (\ref{3x3:simSSA}) as a sum of
other monomials each of which precedes the initial monomial in the
lexicographic ordering. It then follows that the iterations of
(\ref{3x3:simAAA}) and (\ref{3x3:simSSA}) finally give an
expression without any one of the factors $A^3$, $S^3$, $SA^2$ and
$S^2A$. Applying (\ref{3x3:simAAA}) and (\ref{3x3:simSSA}) to a
polynomial of the third degree gives:
\begin{equation}\label{3x3:simP3}
  P_3(A,S) \sim L(A^2S, ASA, AS^2, SAS),
\end{equation}
where $L$ is a linear combination of the arguments. Moreover, from
relation (\ref{3x3:simP3}) it follows that
$\trace{P_3(A,S)}=L(\trace{A^2},\trace{A^2S},\trace{S^2},\trace{S^3})$.

Applying (\ref{3x3:simAAA}) and (\ref{3x3:simSSA}) to a polynomial
of the fourth degree gives
\begin{equation}\label{3x3:simP41}
  P_4(A,S) \sim L(A^2SA, A^2S^2, ASAS, SASA, SAS^2),
\end{equation}
but the expression allows further simplification. The reason is
the existence of another relations besides
(\ref{3x3:chiA})--(\ref{3x3:chiASS}). For example, none of the
relations (\ref{3x3:simAAA}), (\ref{3x3:simSSA}) can be applied to
the monomial $ASASAS$, but from the Hamilton--Cayley identity for
the matrix $AS$ it follows that $(AS)^3 \sim 0$. Applying
(\ref{3x3:chiA})--(\ref{3x3:chiASS}) to the coefficient of $\alpha
\beta \gamma$ in the identity $\chi_3(\alpha A + \beta S + \gamma
AS)=0$ gives the equation
\begin{equation}\label{3x3:chiAASS}
  A^2S^2 -(SA)^2 -A^2\trace{S^2}/2 -S^2\trace{A^2}/2
  +I\trace{A^2S^2}+I\trace{(AS)^2}=0.
\end{equation}
Neglecting the polynomials of lesser degrees results in
\begin{equation}\label{3x3:simSASA}
  SASA \sim A^2S^2
\end{equation}
that turns (\ref{3x3:simP41}) into  $ P_4(A,S) \sim L(A^2SA,
A^2S^2, ASAS, SAS^2)$.  Taking the trace of (\ref{3x3:chiAASS})
gives the equation that relates $\trace{A^2S^2}$ and
$\trace{(AS)^2}$:
\begin{equation}\label{3x3:trAASS}
   4 \trace{A^2S^2} + 2 \trace{(AS)^2} - \trace{A^2}\trace{S^2} = 0.
\end{equation}

Thus we have found all possible relations for the fourth-degree
monomials. From (\ref{3x3:chiAASS}) and (\ref{3x3:trAASS}) the
necessity follows to add one more invariant to $\Omega_3(A,S)$
--- $\trace{A^2S^2}$, $\trace{(AS)^2}$ or their linear combination
independent on (\ref{3x3:trAASS}), for example, the invariant
$\trace{\left[A,S\right]^2}$ with the commutator
$\left[A,S\right]\equiv AS-SA$.

For the polynomials of the fifth degree, applying of
(\ref{3x3:simAAA}), (\ref{3x3:simSSA}) and (\ref{3x3:simSASA})
gives $ P_5(A,S)\sim L(A^2SAS,ASAS^2)$. To find relations for the
fifth degree monomials we equate to zero the coefficient of
$\alpha\beta^2$ in the identity $\chi_3(\alpha S + \beta AS) = 0$
after simplification that results in $I\trace{ASAS^2} +
I\trace{A^2} \trace{S^3}/6 =0$. Consequently, there are no new
invariants to add to the $\Omega_3(A,S)$.

Using (\ref{3x3:simAAA}), (\ref{3x3:simSSA}) and
(\ref{3x3:simSASA}) for the polynomials of the sixth degree gives
the relation $ P_6(A,S)\sim L(A^2SAS^2)$, specifically, it follows
$\left[A,S\right]^3\sim -6A^2SAS^2$. Analyzing the identity
$\chi_3(\left[A,S\right])=0$ we conclude that
$\left[A,S\right]^3\sim 0$. Another invariant
$\trace{\left[A,S\right]^3}=-6\trace{A^2SAS^2}$ is added to the
set $\Omega_3$.

Finally, $P_6(A,S) \sim 0$, and hence any polynomial of the
matrices $A$ and $S$ can be represented as a polynomial of the
degree not higher than five with coefficients which are
polynomials of $\Omega_3(A,S)$. For the trace of each polynomial
we have $\trace{P(A,S)} = P(\Omega_3)$. The complete minimal set
of invariants of the traceless $3\times3$ matrix (\ref{notation})
is
\begin{equation}\label{Omega3}
    \Omega_3(A,S)=\{ \trace{A^2}, \trace{S^2},
    \trace{A^2S}, \trace{S^3}, \trace{\left[A,S\right]^2},
    \trace{\left[A,S\right]^3}
    \}.
\end{equation}
To construct the basis of the family $\Phi^{(n)}$ we need to take
into account additional relation among invariants in
(\ref{Omega3}). Namely, that the last invariant is similar to
pseudoscalar and its square is expressed as the polynomial on the
others invariants. To prove that we rewrite (\ref{3x3:chiA}) in
the form $I\trace{M^3}/3 = M^3 - M \trace{M^2}/2$ substituting
$M=\left[A,S\right]$. After squaring and applying of the
(\ref{3x3:chiAAS}), (\ref{3x3:chiASS}) and (\ref{3x3:chiAASS}) to
the right hand side we obtain the desired relation. Thus for $d=3$
any operator from $\Phi^{(n)}$  can be decomposed in the basis
\begin{equation}\label{basis3}
\eqalign{ \trace{A^2}^{n_1} \trace{S^2}^{n_2}
    \trace{A^2S}^{n_3} \trace{S^3}^{n_4}
    \trace{\left[A,S\right]^2}^{n_5}\trace{\left[A,S\right]^3}^{n_6},
    \cr
    n=2n_1+2n_2+3n_3+3n_4+4n_5+6n_6, \quad n_6\leq1.}
\end{equation}

\section{Conclusion}
The leading contribution to the inertial-range behavior of the
structure functions of the vector admixture $\varphi$ has the form
$S_{n}(r)\sim r^{n(1-\eps/2)}(r/L)^{-\xi_{n}}$, where the
anomalous exponents $\xi_n$ are given by eigenvalues of the matrix
of critical dimensions of the family $\Phi^{(n)}$ of scalar
operators of the form $(\pd\varphi)^n$, see \cite{Runov}. The main
results of the paper are finding of the minimal sets $\Omega_d$ of
the invariants of the tensor $\pd_i\varphi_j$ for the spatial
dimensions $d=2$ (\ref{Omega2}) and $d=3$ (\ref{Omega3}). The
method proposed for invariants searching is based on the
Hamilton--Cayley identity and can be generalized to searching of
the joint invariants of the several tensors for arbitrary spatial
dimension. Using the invariants from $\Omega_d$ the bases of the
family $\Phi^{(n)}$ were constructed for arbitrary $n$ in two
dimensions (\ref{basis2}) and three dimensions (\ref{basis3}).

For $d=2$, the basis (\ref{basis2}) has clear physical sense as
$\trace{S^2}\propto\Phi_{\rm dis}$ --- the dissipation operator
and $\trace{A^2}\propto|{\rm rot}\varphi|^2$ --- the enstrophy
operator. The matrix of the critical dimensions become triangular
in the basis. Just as in the scalar admixture case the inertial
range behaviour of the structure function is determined by the
power of the dissipation operator. So the basis (\ref{basis2})
provides the possibility to solve problem in general. With another
representation of the basis (\ref{basis2}) it was obtained
$\xi_{n}=\eps n(n-2)/8+O(\eps^2)$ in \cite{TMF03} that coincides
with the exponents (\ref{dz1e}) of the scalar case.

For $d=3$, the matrix of critical dimensions in the basis
(\ref{basis3}) has a general form. However, taking into account
the dependencies between the operators radically reduces the size
of the matrix. For example, the family $\Phi^{(18)}$ contains
$47246$ operators for general $d$, but only $154$ of them appear
independent for $d=3$. This gives the possibility to calculate the
anomalous exponents (in the linear approximation on $\eps$):
$\xi_4\approx 0.546\eps$, $\xi_6\approx 1.75\eps$, $\xi_8\approx
3.66\eps$, $\xi_{10}\approx 6.27\eps$, $\xi_{12}\approx 9.58\eps$,
$\xi_{14}\approx 13.6\eps$, $\xi_{16}\approx 18.3\eps$,
$\xi_{18}\approx 23.7\eps$.

\vspace{5mm}

\section*{Acknowledgements}
The author thanks L.~Ts.~Adzhemyan for discussions. The work was
supported by the Russian Foundation for Basic Research (Grant No.
05-02-17524).

\end{document}